\documentclass[a4paper]{jpconf}
\usepackage{graphicx}
\usepackage{mathbbol}
\usepackage{subfigure}
\usepackage{xcolor}
\usepackage{amsmath}
\def\be{\begin{equation}}
\def\ee{\end{equation}}
\def\beq{\begin{equation}}
\def\eeq{\end{equation}}
\def\bc{\begin{center}}
\def\ec{\end{center}}
\def\bea{\begin{eqnarray}}
\def\eea{\end{eqnarray}}
\def\dd{\displaystyle}
\def\nn{\nonumber}
\begin{document}
\title{Pieces of the Flavour Puzzle}
\author{Ferruccio Feruglio}
\address{Dipartimento di Fisica e Astronomia `G.~Galilei', Universit\`a di Padova,\\ Via Marzolo~8, I-35131 Padua, Italy
\vskip .1cm
INFN, Sezione di Padova, Via Marzolo~8, I-35131 Padua, Italy
}

\ead{feruglio@pd.infn.it}

\begin{abstract}
An overview of the flavour problem is presented, with emphasis on the theoretical efforts to find a satisfactory description of fermion masses and mixing angles.
\end{abstract}
\section{Introduction}
The origin of the parameters in the flavour sector of the Standard Model (SM), minimally extended to include massive neutrinos, is one of the most enigmatic questions in particle physics.
Out of the 22 (20 if B-L is conserved) independent low-energy parameters ${\cal Y}_i$, which with some abuse of language can be called Yukawa couplings, 18 have been measured.
Of the remaining four parameters, the absolute scale of neutrino masses is constrained in a limited range, the leptonic Dirac CP-violating phase starts to be constrained by global fits while the two possible 
Majorana phases are still unknown. 
A considerable effort has been devoted to search for a  more economic description, perhaps related to a new principle, such as the gauge principle. Gauge invariance and renormalizability allow to describe strong and electroweak interactions of three copies of fifteen different fermion species in terms of only
three parameters. Nothing similar exists so far in the flavour sector and we usually refer to this as the flavour puzzle.
Another aspect of flavour is related to the new particle threshold around the TeV scale predicted  by all SM extensions addressing the hierarchy problem. Once new TeV particles transforming non-trivially in flavour space are introduced, it is very difficult
to maintain the almost perfect agreement between predictions and observations that reigns in the SM. New sources of flavour-changing neutral currents (FCNC) and CP violations appear and the task is to keep them at an acceptable level.
This is what we commonly mean by flavour problem, to distinguish it from the flavour puzzle.
In this short review the focus will be on the first aspect, I will comment only shortly on the second one. Also, I'm not aiming at reviewing all the existing models, but rather at revisiting some main ideas, guided by my
own prejudices. Several aspects that have been left aside or only briefly mentioned in this paper are covered in a number of reviews \cite{Weinberg:1977hb,Peccei:1997mz,Fritzsch:1999ee,Xing:2014sja}. In particular I will not do justice to the vast literature of fermion masses in grand unified theory \cite{Raby:1995uv,Berezhiani:1995tr,Ross:2000fn}, nor to that discussing the flavour puzzle in the framework of string theory \cite{Chamoun:2005fba,Ibanez:2012xs}.

There are different approaches to the flavour puzzle, with many intermediate possibilities.
We may take a reductionist perspective: the Yukawa couplings ${\cal Y}_i$ should be deduced from first principles. We postulate the existence of a fundamental theory from which ${\cal Y}_i$ can be uniquely determined. Either by proceeding directly from the candidate theory or by appealing to some symmetry or dynamical principle, 
${\cal Y}_i$ are then computed in terms of a small set of input parameters. 
Probably the most striking fact about this program is that nothing approaching a standard theory of ${\cal Y}_i$ exists, despite the decades of experimental progress and theoretical efforts.
In another approach a major role is played by chance. There are many variants and practical implementations of this strategy. The Yukawa couplings ${\cal Y}_i$ are typically mapped to a large number of 
order-one parameters that are considered as irreducible unknowns, like in models with Froggatt-Nielsen abelian flavour symmetries or with fermions living in extra dimensions. 
Also the simplest version of partial compositeness falls into this class. By scanning the order-one parameters we get probability distributions for masses and mixing angles.
Alternatively we start from a fundamental theory, like string theory,  which possesses a vast landscape of solutions, with no privileged ground state. 
The observed Yukawa couplings become environmental quantities and cannot be predicted, like the relative sizes of the solar planetary orbits \cite{Schellekens:2013bpa}. 
We are allowed to ask much less ambitious questions. For instance, if we have knowledge of the statistical distribution of ${\cal Y}_i$ in an hypothetical multiverse where the laws of physics follow our fundamental theory, we can ask how typical are the Yukawa
couplings that we observe. Conversely, barring anthropic selections, we might assume that the observed ${\cal Y}_i$ are typical and try to deduce information on the statistical distribution of ${\cal Y}_i$ in the 
multiverse \cite{Donoghue:2005cf,Hall:2007zj}. Such a variety of open possibilities shows how far we are from the solution of the puzzle and even from identifying the most relevant questions to be addressed.

Most of the parameters ${\cal Y}_i$ are dimensionless and in a dynamical theory of flavour we have essentially no clue about the characteristic scale $\Lambda_f$. If active neutrinos are Majorana particles and B-L
gets violated at a scale $\Lambda$, then $\sqrt{\Delta m^2_{atm}}\approx 0.05$ eV strongly suggests a very large $\Lambda$.
However in general $\Lambda$ and  $\Lambda_f$ are independent from each other. Thus there is no clear relation between $\Lambda_f$ and other possible particle physics thresholds
such as the TeV scale, relevant to the gauge hierarchy problem, or the grand unified scale. This makes it more difficult to identify unambiguous signatures to confirm or rule out a given model 
of fermion masses and mixing angles. For instance, the extrapolation of the Yukawa couplings from the scale $\Lambda_f$ down to low-energies where they are measured
can involve new particle threshold and/or unknown parameters, thus affecting our ability to test the high-energy theory.
\section{Lessons from the quark sector}
A first useful observation is that ratios of charged fermion  masses and quark mixing angles can be represented
by powers of the Cabibbo angle. Using $\lambda=0.22$ we have
 \bea
\dd\frac{m_e}{m_\tau}\approx \lambda^{5.4}& \dd\frac{m_d}{m_b}\approx \lambda^{4.3}& \dd\frac{m_u}{m_t}\approx \lambda^{7.4}\\
\dd\frac{m_\mu}{m_\tau}\approx \lambda^{1.9}& \dd\frac{m_s}{m_b}\approx \lambda^{2.3}& \dd\frac{m_c}{m_t}\approx \lambda^{3.6}~~~,
\eea
where all masses have been renormalised at the scale $m_Z$. It is well-known that also the elements of the Cabibbo-Kobayashi-Maskawa  (CKM) mixing matrix can be expressed in terms of powers of $\lambda$:
\be
|V_{ud}|\approx 1~~~~~|V_{us}|\approx\lambda~~~~~|V_{cb}|\approx\lambda^2~~~~~|V_{ub}|\approx\lambda^4\div\lambda^3~~~~~.
\ee
For comparison, in the lepton sector, where the Pontecorvo-Maki-Nakagawa-Sakata (PMNS) mixing matrix is denoted by $U$, we have all $|U_{fi}|$ of order one, except for $|U_{e3}|$ which is of order $\lambda$. The ratio between the solar and the atmospheric neutrino squared mass differences
$\Delta m^2_{21}/|\Delta m^2_{31}|$ is of order $\lambda^2$. Focussing on the quark sector, in a pioneering work \cite{Froggatt:1978nt} Froggatt and Nielsen observed that all the small dimensionless
parameters of the quark sector such as the quark mass ratios and the CKM mixing angles can be interpreted as powers of the breaking parameter
of a flavour symmetry. In this case the flavour symmetry group $G_f$ is abelian, $G_f=U(1)_{FN}$.
A scalar field $\varphi$, carrying by convention a negative unit of the abelian charge $FN$, develops a vacuum expectation value (VEV) that can be parametrized as
\be
\lambda=\langle \varphi\rangle/\Lambda_f<1~~~~~~~~~~~~~FN(\varphi)=-1 ~~~.
\ee
Quarks  carry non-negative ${\rm U(1)}_{FN}$ charges (the case with charges of both signs can be discussed as well)
\be
FN(X_i)\ge 0~~~~~~~~~(X_i=q_i,u^c_i,d^c_i)~~~.
\ee
Under these assumptions the quark Yukawa couplings $y_{u,d}$ 
are given by:
\be
y_u=F_{u^c} Y_u F_q~~~,~~~~~~~~~~y_d=F_{d^c} Y_d F_q~~~,
\label{yud}
\ee
where $Y_{u,d}$ are complex matrices with entries of order one, undetermined by the ${\rm U(1)}_{FN}$ symmetry, while $F_X$ are real diagonal matrices, completely specified  in terms of $\lambda$ by
the charges $FN(X_i)$:
\be
F_X=
\left(
\begin{array}{ccc}
\lambda^{FN(X_1)}&0&0\\
0&\lambda^{FN(X_2)} &0\\
0&0&\lambda^{FN(X_3)}
\end{array}
\right)~~~~~~~~~~~~~~~~~~~~~~~(X_i=q_i,u^c_i,d^c_i)~~~.
\ee
The small quark mass ratios and quark mixing angles originate from the hierarchical structure of the matrices $F_X$. Indeed, by taking $FN(q_1)>FN(q_2)>FN(q_3)\ge 0$ we get
\be
(V_{u,d})_{ij}\approx \frac{F_{q_i}}{F_{q_j}}<1~~~(i<j)
\ee
for the matrices $V_{u,d}$ defining the CKM mixing matrix $V_{CKM}=V_u^\dagger V_d$. Independently from the specific charge choice, this framework predicts
\be
V_{ud}\approx V_{cs}\approx V_{tb}\approx O(1)~~~~~~~~~~~~~~~~~~~~V_{ub}\approx V_{td}\approx V_{us}\times V_{cb}~~~,
\label{Vij}
\ee
the last equality being correct within a factor of two. With $\lambda\approx 0.2$, the correct order of magnitudes of the $V_{CKM}$ matrix elements can be reproduced by choosing, for instance,
$FN(q)=(3,2,0)$. The correct order of magnitudes of the quark mass ratios can be reproduced by choosing, for example
\be
FN(q)=(3,2,0)~~~~~~~~~~FN(u^c)=(4,2,0)~~~~~~~~~~FN(d^c)=(1+r,r,r)~~~,
\label{qcharges}
\ee
$r$ being a non-negative integer. If there is only one Higgs doublet, then we need $r$ close to 2 to match the ratio $m_t/m_b$. If two Higgs doublets are present, other choices are possible by varying $\tan\beta=v_u/v_d$. 
Several aspects of this class of models have been discussed in refs. \cite{Leurer:1992wg,Leurer:1993gy,Ibanez:1994ig,Binetruy:1994ru,Binetruy:1996xk,Irges:1998ax,Dudas:1995yu,Dudas:1996fe}.

The construction relies on a spontaneously broken abelian flavour symmetry, but 
the final results (\ref{yud}-\ref{Vij}) are valid in a more general context, where no symmetry is present to start with. 
A simple example is provided by a model with an extra spatial dimension, compactified on an orbifold $S^1/Z_2$ to allow for 4D chiral fermions.
The Lagrangian for a 5D spinor $\Psi(x,y)$ reads:
\bea
{\cal L}&=&i\overline{\Psi}\Gamma^M D_M \Psi+m \overline{\Psi} \Psi\nn\\
&=&i\overline{\Psi}\gamma^\mu \partial_\mu \Psi-\overline{\Psi}\gamma_5 \partial_y \Psi+m \overline{\Psi} \Psi+...
\eea
where the mass $m$ should be odd under the $Z_2$ parity sending $y$ into $-y$. A possible choice is
\be
m=M\epsilon(y)~~~,
\ee
$M$ being a real constant and $\epsilon(y)$ the periodic sign function. The 5D spinor has left (L) and right (R)  chiralities in four dimensions
\be
\Psi=
\left(
\begin{array}{c}
\Psi_L\\
\Psi_R
\end{array}
\right)
\ee
with opposite $Z_2$ parities, such that only the even component developes a massless (zero) mode.
Choosing, for instance, $\Psi_L$ even and $\Psi_R$ odd, the equation satisfied by the zero mode of $\Psi_L$ is:
\be
\partial_y \Psi_L^0+M\epsilon(y)\Psi_L^0=0~~~.
\ee
The solution has an exponential dependence on $y$
\be
\Psi^0_L(x,y)=\sqrt{\frac{2M}{1-e^{-2M\pi R}}}e^{-M|y|}\psi(x)~~~,
\ee
where the first factor provides the correct normalization. The zero mode is localized near $y=0(\pi R)$ for $M>0(<0)$. In the limit $M=0$
the zero mode becomes flat in $y$. A formally identical solution holds for the zero mode of $\Psi_R$, if we choose $\Psi_L$ odd and $\Psi_R$ even
and we start from a 5D mass term with the opposite sign. If the Higgs field is strictly localized at one of the two branes, for instance the one at $y=0$,
the Yukawa interactions will be proportional to a Dirac delta $\delta(y)$ and we can reproduce the same pattern of Yukawa coupling of eq. (\ref{yud}) with matrices $F_{X}$ now given by \cite{Kaplan:2001ga}
\be
F_{X_i}=\sqrt{\frac{2\mu_i}{1-e^{-2\mu_i\rho}}}~~~,
\label{wf}
\ee
where $\mu_i$ and $\rho$ are specified in terms of the bulk quark masses and the geometry of the extra dimension, see table \ref{t1}.
\begin{table}[h!]
\caption{\label{t1} Parameters $\mu_i$ and $\rho$ in models with an extra dimension compactified on an interval. In a flat (warped) metric the Higgs field is localized on the brane $y=0$ ($y=R'$) and the ultraviolet cut-off
is denoted by $\Lambda$ ($1/R$). The fermions are described by five-dimensional spinors, with bulk masses $M_i$. In the warped case, when the framework is applied to the gauge hierarchy problem \cite{Randall:1999ee}, $R,R'$ are length scales of the order of the inverse Planck mass and the inverse TeV scale, respectively. } 
\begin{center}
\begin{tabular}{lcc}
\br
ED& $\mu_i$& $\rho$\\
\mr
Flat ~~~~~$[0,\pi R]$& $M_i/\Lambda$& $\Lambda \pi R$\\
Warped $[R, R']$& $1/2-M_i R$& $\log R'/R$\\
\br
\end{tabular}
\end{center}
\end{table}
The suppression factors $F_{X_i}$ represent the values that the profiles of the fermion zero modes
take at the brane where the Higgs field is localized. On that brane, generic $O(1)$ Yukawa couplings $Y_{u,d}$ with the bulk quark fields are postulated.
The role of the Froggatt-Nielsen charges is here played by $\mu_i$ and $\rho$ that determine the profiles along the extra dimension of the zero-mode wave functions:
\be
\sqrt{\rho}F_{X_i}=\sqrt{\frac{\xi_i}{1-e^{-\xi_i}}}
\approx
\left\{
\begin{array}{ll}
\sqrt{\xi_i}& \xi_i\gg1\\
1&|\xi_i|\ll1\\
\sqrt{-\xi_i}~e^{\xi_i/2}& \xi_i\ll-1\\
\end{array}~~~~~~~~~~~~~~~~(\xi_i=2\mu_i \rho)~~~.
\right.
\ee
There is no flavour symmetry: the hierarchical structure of quark masses and mixing angles is dictated by geometry in the compact space.

Similarly, in the partial compositeness
scenario \cite{Kaplan:1991dc}, light fermions get hierarchical masses from the mixing between an elementary sector and a composite one. 
As a toy realization of this idea, consider a model where
the composite sector contains, for each SM fermion, a pair of heavy fermions allowing a Dirac mass term of the order of the compositeness scale and a 
mixing term with the SM fields \cite{Contino:2006nn,Buras:2011ph}
\bea
{\cal L}_Y&=&-u^c \Delta_u U-d^c \Delta_u D-Q^c \Delta_q q\nn\\
&&-U^c M_u U-D^c M_d D-Q^c M_q Q\nn\\
&&-U^c Y_u (\tilde{\Phi}^\dagger Q)-D^c Y_d (\Phi^\dagger Q)-(Q^c \tilde{\Phi}) \tilde{Y}_u U-(Q^c \Phi) \tilde{Y}_d D+h.c.
\eea
The first line represents the mixing between elementary and composite sector, the second line displays Dirac mass terms for the fermions of the composite
sector and the third line shows the Yukawa interactions that, by assumption, are restricted to the composite sector alone and described by strong couplings
$Y_{u,d},  \tilde{Y}_{u,d}\ge 1$. By integrating out the composite sector under the assumption $M_i\gg v$, we get low-energy Yukawa interactions for the elementary sector
whose leading order (LO) terms have the structure given in eq. (\ref{yud}) with matrices $F_{X}$ parametrizing the elementary-composite mixing:
\be
F_{u^c}=\Delta_u M_u^{-1}~~~,~~~~~~~F_{d^c}=\Delta_d M_d^{-1}~~~,~~~~~~~F_q= M_q^{-1}\Delta_q~~~.
\ee

The same pattern arises when matter chiral multiplets $X_i$ of the MSSM are coupled to a superconformal sector in some finite energy range \cite{Nelson:2000sn,Poland:2009yb,Craig:2010ip},
from an ultraviolet (UV) scale $\Lambda$ down to a lower scale $\Lambda_c$. Generic $O(1)$ Yukawa couplings $Y_{ij}$ at the scale $\Lambda$
\be
w=X_i Y_{ij} X_j H+....
\ee
can undergo a sizable renormalization induced by the corrections to the Kahler potential. In the superconformal window the chiral multiplet $X_i$ can
have a large positive anomalous dimension $\gamma_i$ and the Kahler potential at the scale $\Lambda_c$ becomes
\be
K=\sum_i Z_i(\Lambda_c) X_i^\dagger X_i+...
\ee
where
\be
Z_i(\Lambda_c)=Z_i(\Lambda)\left(\frac{\Lambda_c}{\Lambda} \right)^{-\gamma_i}~~~~~~~~~~~~~~~~~Z_i(\Lambda)\approx 1
\ee
Moving to a basis of canonical kinetic terms, the Yukawa couplings at the scale $\Lambda_c$ are renormalized
\be
Y_{ij}(\Lambda_c)= F_{X_i} Y_{ij} F_{X_j} ~~~~~~~~~~~~~~~~~~~~~~~~~~~~F_{X_i}=\left(\frac{\Lambda_c}{\Lambda} \right)^{\frac{\gamma_i}{2}}<1
\ee
and we find again the same pattern of eq. (\ref{yud}), without imposing any symmetry.

In the previous examples the anarchical pattern of $Y_{u,d}$ may result in strong bounds on the scale of new physics $\Lambda_{NP}$ associated to particles carrying flavour quantum numbers
and representing new sources of FCNC and/or CP violation. In the absence of a concrete realisation, it is difficult to estimate reliably the corresponding effects, also because
in general the scale of new physics $\Lambda_{NP}$ and the scale of flavour physics $\Lambda_f$ are independent from each other. A possibility is offered by a spurion analysis 
\cite{Agashe:2004cp}, analogous to that prescribed by Minimal Flavour Violation (MFV)\cite{D'Ambrosio:2002ex}. To this purpose we assume that the new degrees of freedom have non-trivial flavour properties 
and that the flavour-violating effects are completely specified by the same spurions that are responsible for fermion masses and mixing angles. Moreover we assume that the dominant flavour-violating contributions admit an expansion in power series of the spurion fields. 
We start by noticing that the pattern of eq. (\ref{yud}) is compatible with the flavour symmetry $G_f={\rm SU(3)}^3\times {\rm SU(3)_H}^3$ with quarks transforming only under ${\rm SU(3)}^3$ as 
\be
q=(3,1,1)~~~~~~~~~~u^c=(1,\overline{3},1)~~~~~~~~~~d^c=(1,1,\overline{3})~~~.
\label{qmfv}
\ee
The full symmetry $G_f$ is explicitly broken by both the matrices $Y_{u,d}$ and $F_X$. However it can be formally restored by treating $Y_{u,d}$ and $F_X$ as non-dynamical spurion fields possessing
suitable transformation properties. To this aim the Yukawa couplings should transform only under the ``hidden" group
${\rm SU(3)_H}^3$: 
\be
Y_u=(3,\overline{3},1)_H~~~~~~~~~~Y_d=(3,1,\overline{3})_H~~~.
\ee
The suppression matrices $F_X$ are the interface between ${\rm SU(3)_H}^3$ and ${\rm SU(3)}^3$, and they are assigned appropriate transformations
under both factors to guarantee the invariance of the Yukawa interactions described by eq. (\ref{yud}) under ${\rm SU(3)}^3\times {\rm SU(3)_H}^3$.
The starting point of the spurion analysis is similar to that of MFV. Indeed  the maximal flavour symmetry felt by quarks is ${\rm SU(3)}^3$, as in MFV.
However there are more spurions than in MFV, the irreducible ones including now $F_q$, $F_{u^c}$, $F_{d^c}$, $Y_u$ and $Y_d$. One of the most 
dangerous effects originates from the effective operator
\be
\frac{1}{\Lambda_{NP}^2}(\overline{q} F_q^\dagger \gamma_\mu F_q q)(\overline{d^c} F_{d^c}^\dagger \gamma^\mu F_{d^c} d^c)\approx \frac{1}{\Lambda_{NP}^2 \langle Y_d^2\rangle}\frac{2 m_d m_s}{v^2}~(\overline{s} \overline{d^c})(s^c d)+...
\label{estimate}
\ee
$\langle Y_d^2\rangle$ representing an average $O(1)$ coupling.
The contribution of this operator to the CP-violating $\epsilon_K$ parameter is enhanced at the level of both the hadronic matrix element and the QCD corrections and sets one of the most stringent bounds
on the scale of new physics $\Lambda_{NP}$, see Table \ref{tabIs}.
\begin{table}[h]
\begin{center}
\begin{tabular}{|c|c c|c c|} \hline\hline
\rule{0pt}{1.2em}%
Operator &  \multicolumn{2}{c|}{Bounds on $\Lambda_{NP}$~in~TeV~($c_{\rm NP}=1$)} &
\multicolumn{2}{c|}{Bounds on
$c_{\rm NP}$~($\Lambda_{NP}=1$~TeV) }\cr
&   Re& Im & Re & Im \cr  
 \hline $(\bar s_L \gamma^\mu d_L )^2$  &~$9.8 \times 10^{2}$& $1.6 \times 10^{4}$ 
&$9.0 \times 10^{-7}$& $3.4 \times 10^{-9}$\\ 
($\bar s_R\, d_L)(\bar s_L d_R$)   & $1.8 \times 10^{4}$& $3.2 \times 10^{5}$ 
&$6.9 \times 10^{-9}$& $2.6 \times 10^{-11}$ \\ 
 \hline $(\bar c_L \gamma^\mu u_L )^2$  &$1.2 \times 10^{3}$& $2.9 \times 10^{3}$ 
&$5.6 \times 10^{-7}$& $1.0 \times 10^{-7}$ \\ 
($\bar c_R\, u_L)(\bar c_L u_R$)   & $6.2 \times 10^{3}$& $1.5 \times 10^{4}$ 
&$5.7 \times 10^{-8}$& $1.1 \times 10^{-8}$ \\ 
\hline$(\bar b_L \gamma^\mu d_L )^2$    &  $6.6 \times 10^{2}$ & $ 9.3 \times 10^{2}$ 
&  $2.3 \times 10^{-6}$ &
$1.1 \times 10^{-6}$   \\ 
($\bar b_R\, d_L)(\bar b_L d_R)$  &   $  2.5 \times 10^{3}$ & $ 3.6
\times 10^{3}$ &  $ 3.9 \times 10^{-7}$ &   $ 1.9 \times 10^{-7}$ 
    \\
\hline $(\bar b_L \gamma^\mu s_L )^2$    &  $1.4 \times 10^{2}$ &  $  2.5 \times 10^{2}$   &  
 $5.0 \times 10^{-5}$ &   $1.7 \times 10^{-5}$ \\ 
($\bar b_R \,s_L)(\bar b_L s_R)$  &    $ 4.8  \times 10^{2}$ &  $ 8.3  \times 10^{2}$  & 
   $8.8 \times 10^{-6}$ &   $2.9 \times 10^{-6}$  
  \\ \hline\hline
\end{tabular}
\caption{\label{tabIs} Bounds on dimension-six  $\Delta F=2$  
operators, from ref. \cite{Isidori:2010kg,Isidori:2013ez}. The overall coefficient
of the operators is $c_{NP}/\Lambda^2_{NP}$. The operator in eq. (\ref{estimate}) is the one in the second row.}
\end{center}
\end{table}
\noindent
Assuming a generic $O(1)$ phase for the overall coefficient we need
\be
\langle Y_d\rangle~ \Lambda_{NP}>20~{\rm TeV}
\ee
not to spoil the SM prediction for $\epsilon_K$. This, together with other constraints, suggests that a fully anarchical pattern in $Y_{u,d}$ is probably not tenable if new flavoured physics 
is present at the TeV scale \cite{Csaki:2008zd}. 

When such a spurion analysis is applicable, the estimate of eq. (\ref{estimate}) represents a sort of lower bound on the size of the expected effect and larger contributions are possible  \cite{Dudas:2010yh}.
For example in supersymmetric extensions of the SM with a U(1)$_{FN}$ flavour symmetry the operator considered in eq. (\ref{estimate}) receives contributions from box diagrams
with squarks/gluino exchange that are typically larger than the one quoted  in eq. (\ref{estimate}). The reason is that in the U(1)$_{FN}$ case the true flavour symmetry is much weaker
than ${\rm SU(3)_H}^3\times {\rm SU(3)}^3$ and it allows sizable off-diagonal terms in both LL and RR blocks for the first two generations of the down squark mass matrix. For instance, with the charge assignment
of eq. (\ref{qcharges}), the mass insertions $(\delta^d_{12})_{LL}$ and $(\delta^d_{12})_{RR}$ are both proportional to $\lambda$ and the operator $(\overline{s} \overline{d^c})(s^c d)$
has an overall parametric suppression $1/16\pi^2 \times \lambda^2/\Lambda_{NP}^2$, milder than the one in eq. (\ref{estimate}). 

\section{From quarks to leptons}
In the lepton sector we have no evidence for strong hierarchies in mixing angles or in neutrino masses. Hierarchy shows up at the level of charged lepton masses. In terms of the suppression factors $F_{X_i}$
this means
\be
F_{e^c_1}\ll F_{e^c_2}\ll F_{e^c_3}~~~~~~~~{\rm and}~~~~~~~~~F_{l_1}\approx F_{l_2}\approx F_{l_3}~~~.
\label{lept}
\ee
For example an acceptable set of charges is
\be
FN(e^c)=(4,2,0)~~~~~~~~~~FN(l)=(s+t,s,s)~~~~~~~~~~(s\ge0,t=0,1)~~~.
\label{lcharges}
\ee
Here we focus on Majorana neutrinos. In the context of a type I see-saw mechanism right-handed neutrinos $\nu^c$ have their own suppression matrices $F_{\nu^c}$.
Yukawa couplings $y_{\nu,e}$ and the mass matrix $M$ of $\nu^c$ read
\be
y_\nu=F_{\nu^c} Y_\nu F_l~~~,~~~~~~~y_e=F_{e^c} Y_e F_l~~~,~~~~~~~M=F_{\nu^c} Y_c F_{\nu^c} M_0~~~,
\label{Ynue}
\ee
where $Y_{\nu,e,c}$ are complex matrices with unknown entries of order one and $M_0$ is a mass parameter. At low-energy the active neutrino mass matrix $m_\nu$ is given by
\be
m_\nu=-F_l~ (Y_\nu^T  Y_c^{-1} Y_\nu) ~F_l~ v^2/M_0~~~,
\label{mmnu}
\ee
with no dependence on the suppression matrices $F_{\nu^c}$. 

\begin{figure}[!ht]
\centering
\subfigure{\includegraphics[width=0.4\textwidth]{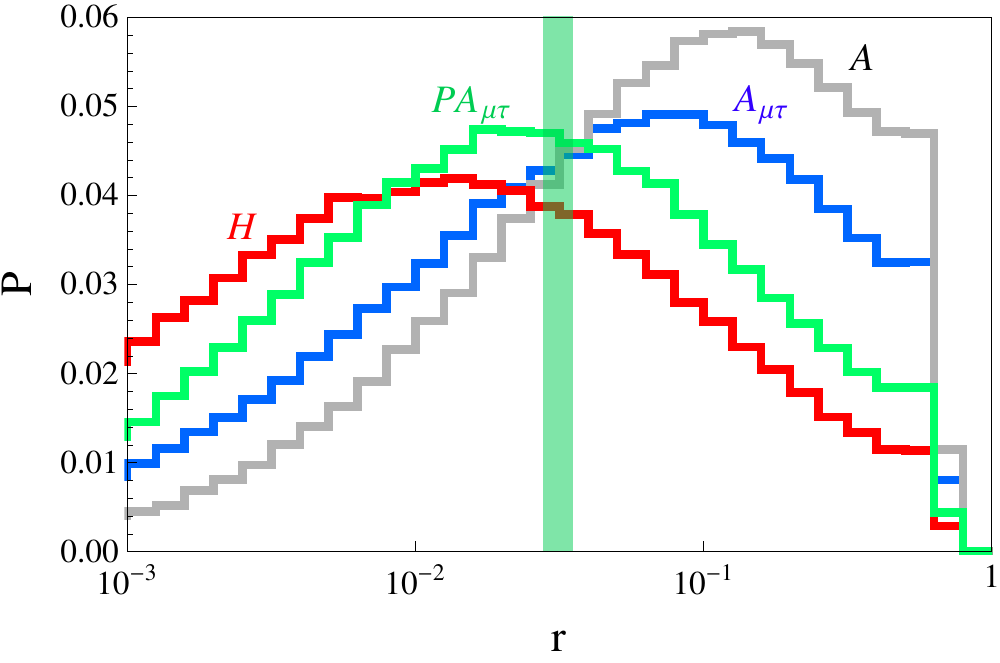}}\quad \hspace*{-0.25cm}
\subfigure{\includegraphics[width=0.4\textwidth]{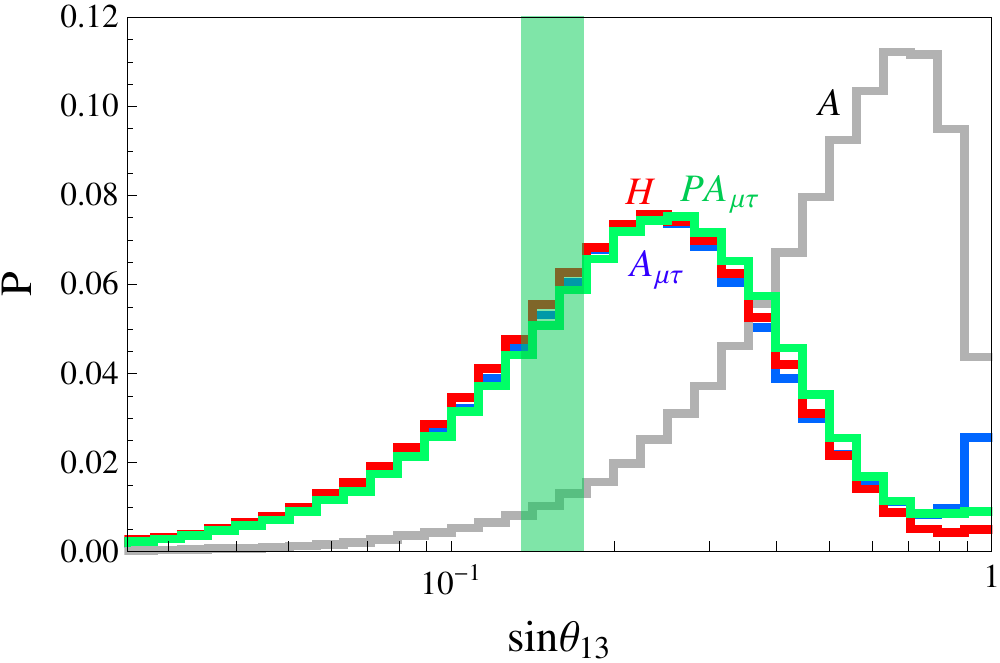}}\quad \\
\subfigure{\includegraphics[width=0.4\textwidth]{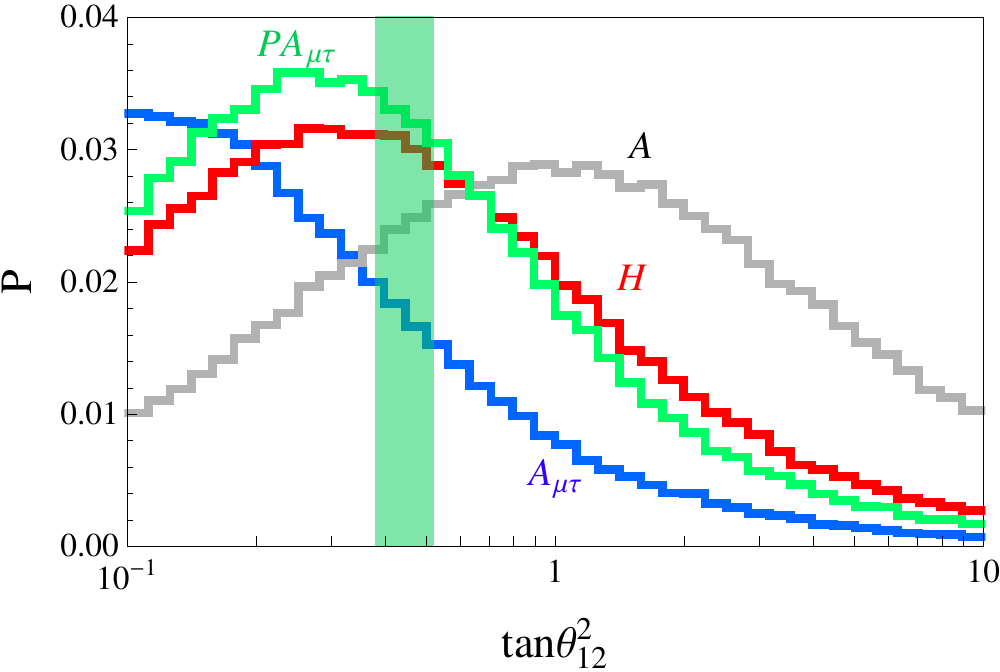}}\quad\hspace*{-0.25cm}
\subfigure{\includegraphics[width=0.4\textwidth]{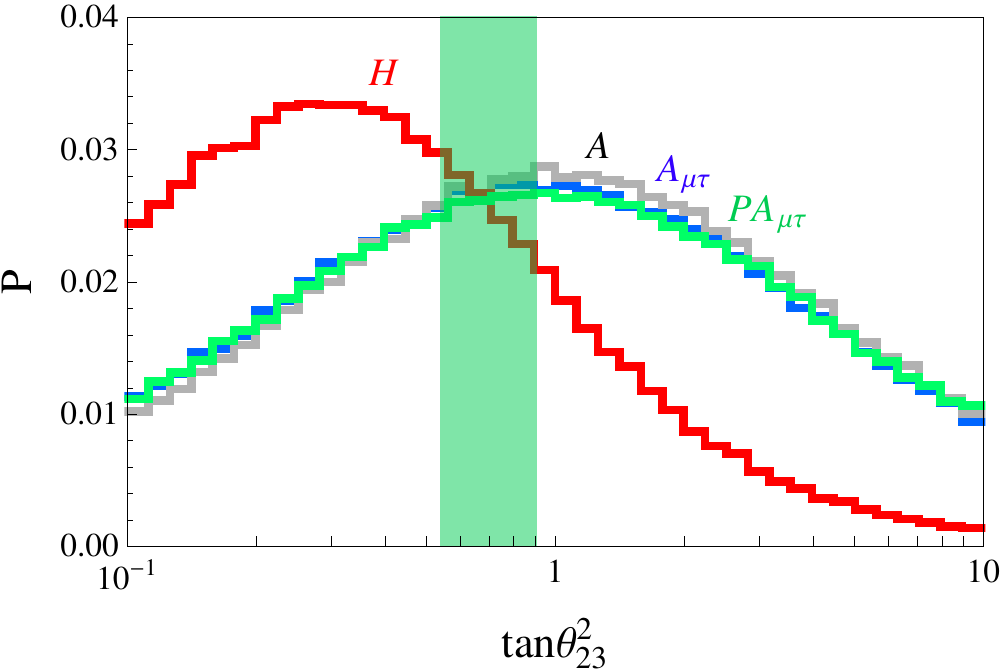}}\quad
\caption{Probability distributions of $r=\Delta m^2_{sol}/\Delta m^2_{atm}, \sin\theta_{13}, \tan^2\theta_{12}, \tan^2 \theta_{23}$, from ref. \cite{u3}, within type I see-saw.
The modulus (argument) of the complex random coefficients has been generated in the interval $[0.5,2]$ ($[0,2\pi]$) with a flat distribution.
For $A$ and $A_{\mu\tau}$, $\lambda=0.2$ has been used, for $H$ and $PA_{\mu \tau}$, $\lambda=0.4$ is taken. The shaded vertical band
emphasizes the experimental $2\sigma$ window.}
\label{figx}
\end{figure}

A drastic realization of this picture is the framework of Anarchy \cite{Hall:1999sn,Haba:2000be,deGouvea:2003xe,Espinosa:2003qz,deGouvea:2012ac}, which corresponds to the case
\be
F_{l_1}= F_{l_2}= F_{l_3}~~~{\rm or}~~~ t=0
\label{anarchy}
\ee
In the anarchic framework the mass matrix for light neutrinos is
\be
m_\nu=
\left(
\begin{array}{ccc}
O(1)&O(1)&O(1)\\
O(1)&O(1)&O(1)\\
O(1)&O(1)&O(1)
\end{array}
\right)~m_0~~~~~~~~~~m_0=\frac{v^2}{M_0}~~~,
\ee
with undetermined order-one matrix elements. This implies mixing angles and neutrino mass ratios of $O(1)$, in rough agreement with the data. No special values for these quantities is expected.
Indeed, before we knew $\theta_{13}$ from the experiments, Anarchy successfully anticipated values close to the upper bound at the time. Global fits of present data hint at deviations of the atmospheric mixing angle 
$\theta_{23}$ from $\pi/4$. Today these indications are still weak, as shown by the instability of the best fit value against different fitting procedures. The persistence of these deviations
in future tests would further strengthen the case for Anarchy. Anarchy represents an extreme possibility
and milder realization of the relations (\ref{anarchy}) are possible. For instance, in the context of SU(5) grand unified models, with a Froggatt-Nielsen ${\rm U(1)}_{FN}$ abelian symmetry, neutrino masses and mixing angles can be reproduced, at the level of order of magnitudes, by several choices of the FN charges for the $\overline{5}$ multiplets hosting the lepton doublets, as shown in table \ref{t2}. $FN$ charges for fermions in the $10$ representations can be suitably chosen so that,
by varying the unknown order-one parameters, reasonable distributions for charged lepton mass ratios, quark mass ratios and quark mixing angles are obtained \cite{u1,u2,u3,u4,u5}.
A naive comparison of the distributions for neutrino masses and mixing angles with data do not appear to favor Anarchy over the other possible charge assignments, as can be seen from fig. \ref{figx}.
\begin{table}[h!]
\caption{\label{t2} Possible choices of $FN$ charges for the $\overline{5}$ representation in a class of SU(5) grand unified models, from ref. \cite{u3}. The second column shows the value of the $FN$ symmetry breaking parameter optimizing the fit to fermion masses and mixing angles.}
\begin{center}
\begin{tabular}{lcc}
\br
&$FN(\overline{5})$& $\lambda$\\
\mr
$A$& (0,0,0)&-\\
$A_{\mu\tau}$& (1,0,0)&0.25\\
$PA_{\mu\tau}$& (2,0,0)&0.35\\
$H$& (2,1,0)&0.45\\
\br
\end{tabular}
\end{center}
\end{table}
I would personally find more appropriate to use the term Anarchy to denote the approaches giving rise to the results (\ref{yud}) and (\ref{Ynue}-\ref{mmnu}) where the absence of any special pattern resides in the matrices $Y$, rather than to indicate the special case defined in eq. (\ref{anarchy}).

If this framework also comprises new flavoured particles at the TeV scale, severe bounds from lepton flavour violation (LFV) apply, under assumptions analogous to those spelled for the quark sector. The irreducible sources of flavour violation in the lepton sector include the matrices
$Y_e$, $F_{e^c}$ and $F_{l}$ and LFV can occur even in the limit of vanishing neutrino masses. Notice that, though MFV cannot be extended unambiguously to the lepton sector \cite{Cirigliano:2005ck}, it predicts
no LFV if neutrinos are massless since in this limit the only relevant spurion in the lepton sector is $Y_e$, which can always be chosen diagonal.
The dipole operator contributing to LFV is
\be
\frac{e}{\Lambda_{NP}^2} e^c \sigma_{\mu\nu} F^{\mu\nu} (F_{e^c}Y_eY_e^\dagger Y_e F_{l}) H^\dagger l~~~.
\ee
The charged lepton mass matrix is proportional to $(F_{e^c}Y_e F_{l})$. In general the combinations  $(F_{e^c}Y_e F_{l})$ and $(F_{e^c}Y_eY_e^\dagger Y_e F_{l})$  are not diagonal in the same basis, 
not even in the case of universal $F_{l}$ of eq. (\ref{anarchy}), and radiative decays of muon and tau are expected. Agreement with the most constraining upper bound, $BR(\mu\to e \gamma)<5.7 \times 10^{-13}$,
requires $\Lambda_{NP}$ well above $10$ TeV \cite{Agashe:2006iy,Csaki:2010aj}. As in the quark sector, a completely anarchical matrix $Y_e$ and flavoured physics at the TeV scale are difficult to reconcile. 
A sufficient condition for the absence of LFV is that $Y_e$, $F_{e^c}$ and $F_{l}$ are diagonal in the same basis, as suggested in some models. Alternatively we can look for special forms of these 
matrices \cite{Chen:2008qg,Perez:2008ee}, possibly dictated by some symmetry requirements.
\section{ Realizations in grand unified theories}
A welcome feature of the above description is that it can be adapted to grand unified theories (GUT) where quarks and leptons are hosted in the same 
multiplet of the gauge group. In SU(5) the gauge symmetry requires
\be
F_q=F_{u^c}=F_{e^c}=F_{10}~~~,~~~~~~~F_l=F_{d^c}=F_{\overline{5}}~~~,~~~~~~~F_{\nu^c}=F_1~~~.
\label{su5}
\ee
Our previous results, eqs (\ref{qcharges}) and (\ref{lcharges}), come very close to this requirement if we choose $r=s$ and $t=1$. If we accept a couple of tunings in the unknown $O(1)$ parameters $Y_{u,d}$,
we can force the equality (\ref{su5}) and still have a decent description of both the quark and lepton mass spectrum.
As we have seen $F_{\nu^c}$ drops from the low-energy quantities. 
It is instructive to consider also the ansatz $F_{\overline{5}}\propto\mathbb{1}$. In this case the hierarchy among fermion masses is entirely due to
$F_{10}$. From eqs. (\ref{yud},\ref{Ynue}) we see that mass ratios in the up-quark sector are the square of the respective mass ratios in the down-quark and in the charged lepton sectors, which is correct
in first approximation. The large lepton mixing corresponds to a large mixing among $d^c$ quarks \cite{Altarelli:1998ns}, unobservable in SM weak interactions, but 
with possible observable effects if transferred from quarks to
squarks in SUSY extensions of the SM \cite{Chang:2002mq}. A minimal model with Higgs bosons in the $5$ representation
would lead to the unrealistic relation $y_e=y_d^T$, but the contributions from other Higgs representations or from non-renormalizable operators can solve this problem \cite{Georgi:1979df,Ellis:1979fg} without altering the picture.

At first sight this description does not seem to be compatible with an SO(10) GUT. The most general renormalizable Yukawa interaction of three copies of fermion generations transforming
as ${\bf 16}$ of SO(10) reads
\be
{\cal L}_Y=-{\bf 16}_i \left[ Y_{10}^{ij}{\bf 10}_H  + Y_{120}^{ij}
 {\bf 120}_H +
Y_{126}^{ij} \overline{\bf 126}_H\right] {\bf 16}_j +h.c.
\label{LY}
\ee
The pattern of Yukawa couplings in eq. (\ref{yud}) can also be thought to arise from a rescaling of the fermions fields, with the constraint
that fermions belonging to a given irreducible representation of the gauge group have to undergo the same renormalization.
By assuming that the matrices $Y_{10}$, $Y_{120}$, $Y_{126}$ have complex elements of order one and that the fields ${\bf 16}$ undergo a wave function renormalization
\be
{\bf 16} \to F_{16} {\bf 16}~~~,
\label{rescale}
\ee
we see that all members of a ${\bf 16}$ representation are affected in the same way. Even accounting for the Clebsch-Gordan coefficients arising from eq. (\ref{LY}) and
the different overall scales associated with the Higgs VEVs $\langle H_{u,d}\rangle$, we cannot reproduce the observed hierarchies of $u$, $d$ and $e$ masses.
Such a discouraging starting point has been successfully modified in a construction by Kitano and Li \cite{Kitano:2003cn}, recently revisited in ref. \cite{Feruglio:2014jla}.

The model is a SUSY SO(10) GUT realized in a flat five-dimensional space time, the fifth dimension being
compactified on an interval $[0,\pi R]$ whose inverse size is of the order of the GUT scale. 
The $N=1$ 5D SUSY corresponds to an $N=2$ 4D SUSY, which is broken down to $N=1$ as a result of 
appropriate boundary conditions. The model comprises a 5D vector supermultiplet decomposing as a 4D vector ${\bf 45_V}$ multiplet plus a 4D chiral ${\bf45_\Phi}$ multiplet. In the bulk there are also three copies of 5D 
hypermultiplets, equivalent to 4D chiral
multiplets ${\bf 16}$ and ${\bf 16^c}$, with bulk masses $M_i$ $(i=1,2,3)$. The boundary conditions allow zero modes only for ${\bf 45_V}$ and ${\bf 16}$. 
A Yukawa superpotential analogous to eq. (\ref{LY}) is localized at the brane $y=0$. Prior to the SO(10) symmetry breaking, the wave functions of fermion zero modes evaluated at $y=0$
effectively drive a rescaling of the Yukawa couplings, as described by (\ref{rescale}). The suppression factors in $F_{16}$ are (see  eq. (\ref{wf})):
\be
F_{16_i}=\sqrt{\frac{2\mu_i}{1-e^{-2\mu_i\rho}}}~~~,
\label{wf2}
\ee
where $\mu_i=M_i/\Lambda$ and $\rho=\Lambda \pi R$. At this stage the Yukawa interactions are not yet able to reproduce the known hierarchies of $u$, $d$ and $e$ masses. The key ingredient of the model resides in
the gauge interaction of the hypermultiplets. The 5D SUSY gauge interaction contains a 4D Yukawa interaction among ${\bf 16}_i$, ${\bf 16^c}_i$ and ${\bf45_\Phi}$, controlled by the gauge coupling $g_5$, that can be combined with the bulk mass term:
\be
-{\bf 16}^c_i \left[M_i-\sqrt{2} g_5  {\bf45_\Phi}\right] {\bf 16}_i~~~.
\label{bulk}
\ee
The chiral multiplet ${\bf45_\Phi}$ has no zero mode but can acquire a non-vanishing VEV, $\langle{\bf45_\Phi}\rangle=v_\Phi^{3/2}$, that breaks SO(10) down to SU(5)$\times$U(1)$_X$. 
The Yukawa interaction of eq. (\ref{bulk}) gives rise to effective SO(10)-breaking bulk masses:
\be
\mu_i^r=\mu_i-Q_X^r k~~~~~~~~~~~~~~~~~~~~k=\sqrt{2}g_5v_\Phi^{3/2}/\Lambda~~~,
\ee
where $Q_X^r$ is the U(1)$_X$ charge of the different SU(5) components inside the ${\bf 16}$ multiplet: $Q_X^r=(-1,+3,-5)$ for $r=(10,\overline{5},1)$. We are back to the SU(5) case, see eq. (\ref{su5}):
\be
F_{r_i}=\sqrt{\frac{2\mu_i^r}{1-e^{-2\mu_i^r\rho}}}~~~~~~~~~r=(10,\overline{5},1)~~~,
\label{wf3}
\ee
but now the profiles $F_{r_i}$ only depend on four free parameters: $\mu_i$ and $k$. Neutrinos are described within a type I see-saw mechanism, as in eq. (\ref{mmnu}), with masses for heavy Majorana neutrinos
originating from the VEV of the SU(5) singlet in the ${\bf \overline{126}}_H$ representation. 
\begin{figure}[!h]
\centering
\subfigure{\includegraphics[width=0.4\textwidth]{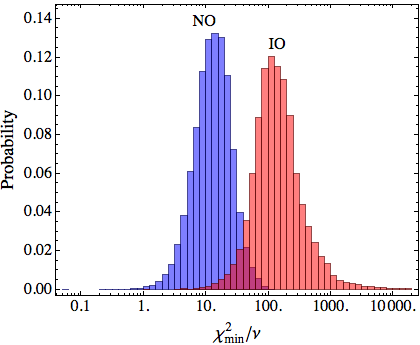}}
\caption{The distributions of minimized $\chi^2/\nu$  for NO and IO in neutrino masses and for
$\tan\beta=50$, from ref. \cite{Feruglio:2014jla}.}
\label{fig2}
\end{figure}

The model contains many parameters of order one. After rephasing of the relevant fields there are 27 real parameters coming
from the matrices $Y_{10}$, $Y_{120}$, $Y_{126}$ and 8 real parameters describing the embedding of the two light Higgs doublets within ${\bf 10}_H$ and ${\bf 120}_H$.
Despite the large number of parameters the agreement with data is not a priori guaranteed, since there are only 4 profile parameters to describe hierarchical mass ratios and mixing angles.
Indeed a fit to an idealized set of 17 observables leads to a good agreement only for large values of $\tan\beta$, for both normal (NO) and inverted (IO) neutrino mass ordering.  
\begin{figure}[!h]
\centering
\subfigure{\includegraphics[width=0.325\textwidth]{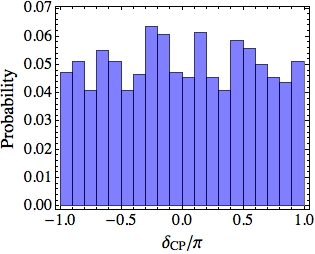}}\quad \hspace*{-0.25cm}
\subfigure{\includegraphics[width=0.31\textwidth]{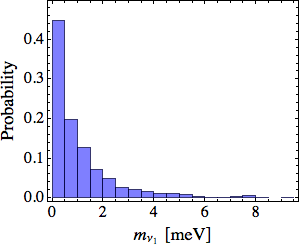}}\quad \hspace*{-0.25cm}
\subfigure{\includegraphics[width=0.32\textwidth]{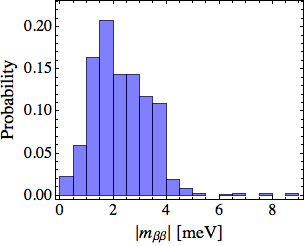}}\quad
\caption{The predictions for various observables obtained for $\chi^2_{\rm min}/\nu<2.21$ in case of
NO and
$\tan\beta=50$, from ref. \cite{Feruglio:2014jla}.}
\label{fig4}
\end{figure}

A closer inspection reveals that fitting fermion masses and mixing angles in the IO case requires a fine-tuning of the Yukawa parameters. By generating a large sample of random order-one
Yukawa parameters, the fit can be repeated  by keeping at each iteration only 12 free parameters, 4 for the profiles and 8 for the relevant Higgs combinations. 
The distributions of the minimum $\chi^2$ over the number of degrees of freedom are shown in fig. \ref{fig2} for NO and IO. 
We see a clear difference between the two cases.  While in the IO case we need about $10^5$ samples to reach a $p$-value close to $0.05$, 
in the NO case in about one percent of the cases we have $p>0.05$. The model needs a severe fine-tuning of the ``anarchical'' parameters in the IO case, 
while the NO one is realized much more naturally. The most probable values of the profile parameters give $F_{\overline{5}}\approx(0.07,0.22,0.63)$, showing that approximate Anarchy is an output rather than an input of the present construction.
\begin{figure}[!h]
\centering
\subfigure{\includegraphics[width=0.45\textwidth]{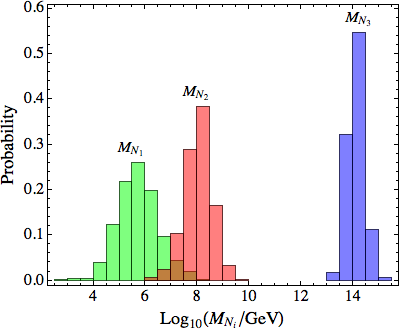}}\quad
\caption{The predictions for the masses of RH neutrinos
obtained for $\chi^2_{\rm min}/\nu<2.21$ in case of NO and $\tan\beta=50$, from ref. \cite{Feruglio:2014jla}.}
\label{fig5}
\end{figure}

Focussing on the NO case, there is no preferred value of the leptonic Dirac CP phase.
The lightest neutrino mass is predicted below $5$ meV, corresponding to a hierarchical neutrino mass
spectrum while $|m_{\beta\beta}|$ lies in the range 0.1-5 meV, see fig. \ref{fig4}. Any positive signal in the
current generation of experiments aiming at measuring neutrino masses or $|m_{\beta\beta}|$ in the
lab would essentially rule out the model. The hierarchy in the right handed neutrino spectrum is
very pronounced and the corresponding mass distributions are peaked around $10^6$ GeV, $10^8$ GeV
and $10^{14}$ GeV, as shown in fig. \ref{fig5}.

In summary, fermion masses and mixing angles are well described by the map in eqs. (\ref{yud},\ref{Ynue},\ref{mmnu}), in terms of input parameters of order one, the elements of the $Y$ matrices. Such
a map can be realized in several different frameworks and does not necessarily need an underlying symmetry. The setup is compatible with both SU(5) and SO(10) grand unification and with the known solution to the gauge hierarchy problem. On the weak side, additional ingredients are probably needed to control the new sources of FCNC and CP-violations arising from new flavoured physics at the TeV scale.
Moreover all entries of the $Y$ matrices are independent free parameters and it is not possible to make absolute predictions, beyond the order-of-magnitude accuracy. 
This is clearly a major limitation, since we would like to test the theory at the level of the best available experimental precision. Finally the  map in eqs. (\ref{yud},\ref{Ynue},\ref{mmnu}) might
be an oversimplified version of a more accurate description and indeed there are several variants of the frameworks briefly mentioned in section 2 that modify the results of eqs. (\ref{yud},\ref{Ynue},\ref{mmnu}).
 
\section{Flavour symmetries}
Given the successful use of symmetries in other contexts of particle physics, it is natural to investigate whether Yukawa couplings can be constrained by a flavour symmetry.
The largest possible classical flavour symmetry of a theory with the particle content of the SM is $G_{MFV}={\rm U(3)}^5$ and corresponds to the limit in which the Yukawa couplings are turned off. The observed fermion masses and mixing angles break $G_{MFV}$ almost completely to a residual symmetry that includes the weak hypercharge and the combination B-L. Similarly, in any realistic model based on flavour symmetries,
the flavour symmetry group $G_f$ is broken. In predictive models the breaking is spontaneous and occurs through the VEVs of a set of scalar fields $\varphi$ transforming non-trivially under $G_f$. The VEVs $\langle \varphi\rangle$ are either postulated or determined by minimizing a $G_f$-invariant energy functional $V(\varphi)$. The Yukawa couplings become dynamical variables
evaluated at the minimum of $V(\varphi)$: ${\cal Y}(\langle\varphi\rangle/\Lambda_f)$. A huge number of models can be constructed according to this set of rules,
depending on the choice of $G_f$ (global, local, continuous, discrete, abelian, non-abelian), and on the choice of representations for scalars and fermions. 

An attempt to start from the full $G_{MFV}$ symmetry is described in refs. \cite{c1,c2,c3,Fong:2013dnk,Maiani:2014vqa}. The framework is that of MFV \cite{D'Ambrosio:2002ex}. 
Under $SU(3)^3$ quarks transform as in eq. (\ref{qmfv}). Yukawa couplings are promoted to spurions transforming as
\be
y_u=(\overline{3},3,1)~~~~~~~~~~y_d=(\overline{3},1,3)~~~,
\ee
to ensure invariance of the Yukawa interactions under $G_{MFV}$. By analyzing a general $G_{MFV}$-invariant potential depending on $y_{u,d}$, it has been proven the existence of stationary points where
\be
y_{u,d}\propto {\tt diag}(0,0,1)~~~~~~~~~~~~~~~~V_{CKM}=\mathbb{1}~~~,
\ee 
pointing to an approximate U(2)$^3$ symmetry of the quark spectrum and providing a good first-order approximation of quark masses and mixing angles. By extending $G_{MFV}$ to the lepton sector, other stationary points have been identified:
\be
y_e\propto {\tt diag}(0,0,1)~~~~~~~~~~~~~~~~~~~~~~~~~~~~m_\nu=U^*_{PMNS}\hat{m}_\nu U^\dagger_{PMNS}
\ee
\be
\hat{m}_\nu={\tt diag}(m_1,m,m)~~~~~~~~~~
U_{PMNS}=\left(
\begin{array}{ccc}
1&0&0\\
0&\frac{1}{\sqrt{2}}&\frac{1}{\sqrt{2}}\\
0&-\frac{1}{\sqrt{2}}&\frac{1}{\sqrt{2}}
\end{array}
\right)\times {\tt diag}(-i,-i,1)~~~.
\ee
Such a solution nicely exhibits maximal $\theta_{23}$, but needs $m_1=m$ to further enforce a large, undetermined $\theta_{12}$.
Goldstone bosons arising from the breaking of $G_{MFV}$ can be eliminated by gauging the flavour symmetry \cite{Grinstein:2010ve}.
Correction terms are needed to promote the leading-order picture into a more realistic theory. 
If there is new physics close to the TeV scale, the advantage of assuming the largest possible flavour symmetry $G_{MFV}$ is the high degree of protection against FCNC induced by the new flavoured degrees of freedom. If the only sources of flavor symmetry breaking are the SM Yukawa couplings, such a maximal symmetry is minimally violated \cite{D'Ambrosio:2002ex}. The classification of the expected effects is unambiguous in the quark sector, and can be extended in several ways in the lepton sector \cite{Cirigliano:2005ck,Joshipura:2009gi}.

Continuous flavour symmetry groups such as SO(3) and SU(3) have been proposed \cite{Ross:2004qn,deMedeirosVarzielas:2005ax,Antusch:2007re}, also in the context of grand unified theories \cite{King:2003rf,King:2005bj}, with the three fermion families assigned to an irreducible triplet representation. 
Charged fermions of the first two generations are much lighter than those of the third generation and consequently within SU(3)/SO(3) we are forced to 
introduce large breaking terms. Alternatively we can start from the smaller flavour group U(2) and assign the first two generations to doublets and the third one to singlets.
Within the simplest realisation, such an assignment in the quark sector leads to \cite{Barbieri:1996ww}
\be
y_{u,d}=
\left(
\begin{array}{ccc}
0& \epsilon'_{u,d}&0\\
\epsilon'_{u,d}& \epsilon_{u,d}&{\cal O}(\epsilon_{u,d})\\
0&{\cal O}(\epsilon_{u,d})&1
\end{array}
\right)(y_{u,d})_{33}~~~,
\label{u2}
\ee 
where phases have been omitted. To correctly reproduce quark masses without appealing to cancellations among the matrix elements we need $|\epsilon'_{u,d}|\ll |\epsilon_{u,d}|\ll 1$, which corresponds to a sequential breaking of U(2):
\be
U(2)\xrightarrow{\epsilon} U(1)\xrightarrow{\epsilon'} {\tt nothing}~~~.
\ee
The following relations can be derived \cite{Gatto:1968ss,Hall:1993ni} by a perturbative diagonalization of $y_{u,d}$:
\bea
|V_{us}|&=&\left| \sqrt{\frac{m_d}{m_s}} -e^{i\varphi}\sqrt{\frac{m_u}{m_c}} \right|
\label{predu1}\\
\left|\frac{V_{ub}}{V_{cb}}\right|&=&\sqrt{\frac{m_u}{m_c}}\label{predu2}\\
\left|\frac{V_{td}}{V_{ts}}\right|&=&\sqrt{\frac{m_d}{m_s}}~~~,\nn
\eea
up to corrections of relative order $\epsilon$, numerically close to few percent. The phase $\varphi$ is arbitrary. These predictions can be translated into constraints on the $(\rho,\eta)$ plane
and compared to the region presently allowed by the data, under the assumption that the SM correctly describes all relevant processes.
Using as inputs quark masses, the maximally allowed values for $\sin 2\beta$ and for $|V_{ub}/V_{cb}|$
derived from (\ref{predu2}) are too small and the above set of Yukawa matrices are now excluded at the 3$\sigma$ level \cite{Roberts:2001zy,Kim:2004ki}
\footnote{The most recent results for $\sin 2\beta$ and $|V_{ub}|$ ($\sin 2\beta=0.695\pm0.023(0.692^{+0.020}_{-0.018})$, $|V_{ub}|=36.3\pm1.2(35.7^{+1.6}_{-1.5})\times 10^{-4}$ \cite{UTfit} (\cite{CKMfitter}))
are compatible with the ones in ref. \cite{Kim:2004ki} ($\sin 2\beta=0.739\pm0.048$, $|V_{ub}|=(35.7\pm3.1)\times 10^{-4}$) and, at the same time, more restrictive.}.
Modifications of the ansatz (\ref{u2}), where either the 13 matrix element is non-vanishing or the element 32 is of the same order of the element 33,
have been proposed to recover agreement with the data  \cite{Roberts:2001zy}. 

This is a nice example of a model of fermion masses based on symmetry requirements and leading to testable predictions. 
In this case the predictions were initially supported by data. Later on the precise measurements at the B-factories and the 
improved theoretical knowledge of both perturbative and non-perturbative QCD effects have considerably reduced the errors
on the CKM matrix elements. Also the knowledge of light quark masses has improved and deviations from (\ref{predu2}) are now significant.
Unfortunately not all present models and ideas can be tested at the same level of accuracy. 

Independently of their dynamical origin, $y_{u,d}$ in eq. (\ref{u2}) are an example of {\em textures}, a simple-minded but predictive approach to the problem of fermion masses and mixing angles
pioneered by Fritzsch \cite{Fritzsch:1977za,Fritzsch:1977vd} and Weinberg \cite{Weinberg:1977hb}. In this approach we assume a privileged flavour basis and some special energy scale
where some of the entries of the matrices $y_f$ vanish. In the quark sector, this can lead to relations between the elements of the CKM mixing matrix and the quark mass ratios
which can be precisely tested. Not all zeros give rise to relations among physical quantities.  By performing weak basis transformations, we can generate zeros in $y_f$
that have no physical implications \cite{Branco:1999nb}. For instance, working with two generations, we can always set to zero the $(1,1)$ entry of both $y_d$ and $y_u$ by performing 
a unitary transformation on the SU(2)-singlets quark fields. This transformation corresponds to a change of basis that has no physical consequences. If, in such a basis, we impose
that $y_d$ and $y_u$ are symmetric matrices, this represents a non-trivial requirement. By further assuming small off-diagonal matrix elements, we get in first approximation the well-known Gatto-Sartori-Tonin relation 
\cite{Gatto:1968ss}, eq. (\ref{predu1}). 
Other unitary transformations, such as permutations, preserves the number of zeros and allow to divide the possible patterns of $y_d$ and $y_u$ into equivalence classes with identical predictions.
It is always possible to enforce texture zeros in arbitrary entries of the fermion mass matrices by means of Abelian symmetries \cite{Grimus:2004hf}.

Special attention has been paid to hermitian textures, $y_{u,d}=y_{u,d}^\dagger$.
One such texture is called $n$-zero if $y_u$ and $y_d$ have a total of $n$ zeroes, the off-diagonal ones counting a half.
A typical texture is
\be
y_f=
\left(
\begin{array}{ccc}
0&A_f&0\\
\overline{A_f}& D_f& B_f\\
0& \overline{B_f}& C_f
\end{array}
\right)~~~~~~~~~~~~~~~~~~~(f=u,d)~~~,
\label{texture}
\ee
where $D_u=D_d=0$ in the original proposal by Fritzsch. The 6-zero Fritzsch texture is ruled out since it predicts a too large $|V_{cb}|$.
More general hermitian 6-zero and 5-zero textures have been analyzed under the assumption of hierarchy among the matrix elements \cite{Ramond:1993kv}. All possible combinations
of hermitian 6-zero and 5-zero textures are ruled out by now \footnote{According to ref. \cite{Ponce:2013nsa}, the five independent hermitian 5-zero textures originally proposed by \cite{Ramond:1993kv} are still viable. However the values of $\beta$ 
obtained from these textures in ref. \cite{Ponce:2013nsa} are too small, out of many sigmas from the currently allowed experimental range, except for the texture IV where the agreement is within about two sigmas.}, because they cannot reproduce $|V_{cb}|$, with the only exception of  $y_f$ in (\ref{texture}) with $D_u\ne 0$ and $D_d=0$, 
which is marginally compatible with present data  \cite{Mahajan:2009wd}. 
The 4-zero texture in (\ref{texture}) with both $D_u$ and $D_d$ non vanishing
is still allowed, provided the hierarchy in the 23 block is mild \cite{Xing:2003yj,Verma:2009gf}. In particular $|B_f/C_f|$ should be ${\cal O}(0.1)$. In this case cancellations among the matrix elements are needed
to get $m_s/m_b$, $ m_c/m_t$ and $|V_{cb}|$ in the correct range. Notice that such cancellations were excluded by assumption in the case of $y_{u,d}$ of eq. (\ref{u2}), where the hierarchy between quark masses of second and third generations was attributed to the smallness of $\epsilon$.

More general textures have been analyzed in ref. \cite{Ludl:2015lta}. If no symmetry requirement is imposed, there are viable textures that correctly fit the data. However they do not show any predictive 
power with respect to any of the quark masses and mixing parameters. When $y_{u,d}$ are symmetric matrices, there are several independent 4-zero and 5-zero textures that allow to predict some 
of the light quark masses in terms of the mixing angles and the remaining masses.

In the lepton sector, working in the flavour basis where the charged lepton mass matrix is diagonal, we can study textures of the symmetric matrix for light Majorana neutrinos \cite{Frampton:2002yf}. 
Those with three or more zeros are experimentally excluded, while data still allow seven independent 2-zero textures \cite{Fritzsch:2011qv,Meloni:2012sx}. The requirement that zeroes of the neutrino mass matrix 
should be present in the flavour basis can be relaxed and more general textures have been investigated. Both cases of Dirac and Majorana neutrinos have been analyzed and many independent textures 
in agreement with the existing data have been identified. For a systematic study of all possibilities of texture zeros in the charged-lepton and neutrino mass matrices and for a recent review on the subject 
we refer the reader to ref. \cite{Ludl:2014axa} and ref. \cite{Gupta:2012dma}, respectively.

Other choices of $G_f$ have been considered. For instance refs. \cite{Barbieri:2011ci,Barbieri:2012uh,Barbieri:2014tja} study the case of $G_f=U(2)^3$, as an alternative to MFV to suppress FCNC in supersymmetric extensions
of the SM. In that context, the advantage of $G_f=U(2)^3$ is that squarks of the first two generations can be taken significantly heavier than the third generation ones. 
Furthermore the presence of separate $U(2)$ for left and right-handed fields, provides a sufficient protection of flavour-violating effects in the right-handed sector.
\section{Discrete symmetries}
The data from neutrino oscillations before 2012 were supporting flavour symmetries, especially through the indication of a vanishing reactor angle $\theta_{13}$ and a maximal atmospheric
mixing angle $\theta_{23}$, features that are difficult to attribute to an underlying theory based on pure chance. Today we know with accuracy that $\theta_{13}$ is neither vanishing nor particularly small,
its size being comparable to that of the Cabibbo angle. 

Recent global fits \cite{GonzalezGarcia:2012sz,Capozzi:2013csa,Forero:2014bxa} (see table \ref{t5}) favor a deviation of the atmospheric mixing angle from the maximal value by several degrees and, at the same time,
show a preference for a maximal Dirac CP-violating phase $\delta_{CP}$, though the whole range from $0$ to $2\pi$ is still allowed at 3$\sigma$. These two features are closely related, since
they are mainly driven by the direct comparison between $P_{ee}$ measured by reactor experiments, which essentially determines $\theta_{13}$, and the conversion
probability $P_{\mu e}$ measured by MINOS and T2K, which in turn is sensitive to a combination of $\theta_{13}$, $\theta_{23}$ and $\delta_{CP}$. 
There is not yet a clear indication of the octant $\theta_{23}$ falls in. Furthermore the most precise single experimental determination of the atmospheric angle \cite{Abe:2014ugx}, 
$\theta_{23}=0.514^{+0.055}_{-0.056}(0.511^{+0.055}_{-0.055})$ for NO (IO), is still compatible with $\pi/4$. Probably it is too early to conclude that a maximal $\theta_{23}$ is ruled out by data.

There are few known mechanisms to generate a maximal atmospheric angle.
We know that $\theta_{23}$ cannot be made naturally maximal by renormalization group evolution, barring a fine-tuning of the initial conditions or ad-hoc threshold effects \cite{Ellis:1999my,Casas:2003kh,Chankowski:2001mx,Broncano:2004tz}. Moreover, in the context of
flavour symmetries, $\theta_{23}$ is completely determined by breaking effects, if we accept that $e$ and $\mu$ masses vanish in the limit of exact symmetry \cite{Feruglio:2004gu}. 
Indeed, if the masses of $e$, $\mu$ and $\tau$ are all non-vanishing when the symmetry is exact, then large ${\cal O}(1)$ breaking terms are needed to explain the relative hierarchy
among charged fermion masses. Thus it is more natural to assume that at least $m_e$ and $m_\mu$ are proportional to small symmetry breaking parameters. In this case, when the symmetry is exact,
the lepton mixing matrix is determined up to an arbitrary rotation coming from the charged $e\mu$ lepton sector and the atmospheric mixing angle can only be determined when the symmetry breaking
parameters are turned on. We have no examples of a maximal $\theta_{23}$ from the breaking of an abelian symmetry. If a nearly maximal atmospheric angle is not due to pure chance, we are left with broken non-abelian symmetries.

Before the measurement of $\theta_{13}$ a particularly attractive lepton mixing pattern was the tribimaximal one
\be
U_{TB}=
\left(
\begin{array}{ccc}
\frac{2}{\sqrt{6}}&\frac{1}{\sqrt{3}}&0\\
\frac{1}{\sqrt{6}}&-\frac{1}{\sqrt{3}}&\frac{1}{\sqrt{2}}\\
-\frac{1}{\sqrt{6}}&\frac{1}{\sqrt{3}}&\frac{1}{\sqrt{2}}
\end{array}
\right)
\approx
\left(
\begin{array}{ccc}
0.82&0.58&0\\
0.41&-0.58&0.71\\
-0.41&0.58&0.71
\end{array}
\right)~~~.
\ee

\begin{table}[h!]
\caption{\label{t5} Results of global fits to neutrino oscillation data from ref. \cite{GonzalezGarcia:2012sz,Capozzi:2013csa,Forero:2014bxa} for normal [NO] and inverted [IO] ordering.}
\begin{center}
\begin{tabular}{lccc}
\br
& \cite{GonzalezGarcia:2012sz} & \cite{Capozzi:2013csa} & \cite{Forero:2014bxa} \\
\mr
$\sin^2\theta_{23}$~[NO]& $0.451^{+0.026}_{-0.020}$ & $0.437^{+0.033}_{-0.023}$ & $0.567^{+0.032}_{-0.128}$\\
& & &\\
$\sin^2\theta_{23}$~[IO]& $0.580^{+0.024}_{-0.039}$ & $0.455^{+0.139}_{-0.031}$ & $0.573^{+0.025}_{-0.043}$\\
\mr
$\delta_{CP}/\pi$~[NO]&  & $1.39^{+0.38}_{-0.27}$ & $1.34^{+0.64}_{-0.38}$\\
& $1.44^{+0.42}_{-0.38}$& & \\
$\delta_{CP}/\pi$~[IO]&  & $1.31^{+0.29}_{-0.33}$ & $1.48^{+0.34}_{-0.32}$\\
\br
\end{tabular}
\end{center}
\end{table}

The good agreement between TB mixing and pre-2012 data strongly supported the idea that the true mixing matrix could be described in terms of small corrections to a LO 
mixing matrix $U^0_{PMNS}$, which could be derived from symmetry considerations. The simplest way to reproduce the TB mixing pattern is by exploiting discrete flavour symmetries \cite{Altarelli:2010gt,Ishimori:2010au,King:2013eh,Altarelli:2014dca,King:2014nza}.
The theory is invariant under a discrete flavour symmetry $G_f$, broken down in such a way that neutrino and charged lepton sectors have different residual symmetries $G_\nu$ and $G_e$, 
at least in a LO approximation where small effects are neglected. If neutrinos are of Majorana
type, the most general group leaving $m_\nu$ invariant and the individual masses $m_i$ unconstrained is $Z_2\times Z_2$, a finite group. The subgroup $G_e$ can be continuous,
but $G_e$ discrete is the simplest option. We require a sufficiently large $G_e$ to distinguish the three charged leptons. For instance we can choose $G_e=Z_n$ ($n\ge 3$) or $G_e=Z_2\times Z_2$.
Once $G_e$ and $G_\nu$ have been chosen inside $G_f$, the embedding automatically fixes the relative alignment of $m^\dagger_l m_l$ and $m_\nu$ in flavour space.
Lepton masses are unconstrained but $U^0_{PMNS}$ is determined up to Majorana phases and up to permutations of rows and columns. This freedom apart,
this setup predicts the three mixing angles $\theta^0_{ij}$ and the Dirac phase $\delta^0_{CP}$. In most concrete models, where symmetry breaking is
achieved via VEVs of a set of flavons $\varphi$, the LO results are modified by small corrections of order $u=\langle\varphi \rangle/\Lambda_f$
\be
U_{PMNS}=U^0_{PMNS}+O(u)~~~.
\label{PMNS}
\ee
Before 2012, in the specific case $U^0_{PMNS}=U_{TB}$ these corrections were expected to be very small, of the order of few percent \cite{Altarelli:2005yp,Altarelli:2005yx}, 
not to spoil the good agreement in the predicted value of the solar mixing angle. On this basis
the simplest models reproducing $U_{TB}$ at the LO predicted $\theta_{13}$ not larger than few degrees, now proven to be wrong by experiments.
Discrete flavour symmetries can also be extended to quarks and even incorporated in GUTs, but in the existing constructions 
the symmetry has to be badly broken in the quark sector. 
Discrete flavour symmetries are also relevant in the so called indirect models \cite{King:2013eh}. In this case the breaking of $G_f$ leaves no residual
symmetries and the role of the flavour group is mainly to get specific vacuum alignments of the scalar fields that control fermion masses. 

\begin{figure}[!h]
\centering
\subfigure{\includegraphics[width=0.4\textwidth]{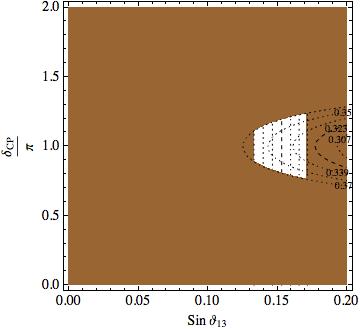}}\quad\quad\quad
\subfigure{\includegraphics[width=0.4\textwidth]{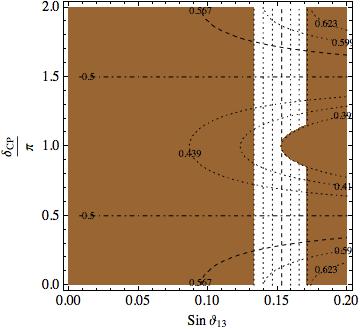}}\quad
\caption{Left panel: contours of equal $\sin^2 \theta_{12}$ in the plane $(\sin\theta_{13},\delta_{CP}/\pi)$, when $U_{PMNS}=U_{12}\times U_{BM}$. The brown region is excluded at $3\sigma$, assuming $[0.0177\div0.0294]$, $[0\div2\pi]$ and
$[0.278\div0.375]$ as $3\sigma$ ranges for $\sin\theta_{13}$, $\delta_{CP}$ and  $\sin^2 \theta_{12}$, respectively. Right panel:
contours of equal $\sin^2 \theta_{23}$ in the plane $(\sin\theta_{13},\delta_{CP}/\pi)$, when $U_{PMNS}=U_{TM_2}$. The brown region is excluded at $3\sigma$, assuming $[0.0177\div0.0294]$, $[0\div2\pi]$ and
$[0.392\div0.643]$ as $3\sigma$ ranges for $\sin\theta_{13}$, $\delta_{CP}$ and  $\sin^2 \theta_{23}$, respectively.}
\label{figctm}
\end{figure}

Several modifications of the simplest models based on discrete symmetries have been proposed to match the most recent data.
If we keep adopting $U^0_{PMNS}=U_{TB}$ as LO approximation, an economic way to reproduce the actual value of $\theta_{13}$ is to introduce large correction terms, $O(u)\approx 0.2$. This is also viable in other schemes where $U^0_{PMNS}$ differs substantially from $U_{TB}$,
such as the so-called bimaximal mixing. 
\be
U_{BM}=
\left(
\begin{array}{ccc}
\frac{1}{\sqrt{2}}&\frac{1}{\sqrt{2}}&0\\
\frac{1}{2}&-\frac{1}{2}&\frac{1}{\sqrt{2}}\\
-\frac{1}{2}&\frac{1}{2}&\frac{1}{\sqrt{2}}
\end{array}
\right)~~~.
\ee
Introducing large corrections has the disadvantage that beyond the LO the number of independent contributions to the mixing matrix is generally quite large. If their typical size
is about 0.2, all mixing angles tend to be affected by generic corrections of this type and predictability is lost. Moreover large correction terms are dangerous if new sources
of flavour changing and/or CP violation are present at the TeV scale. Thus some assumptions about the dominant source of corrections are needed. For example, a reasonable possibility is  
to perturb the BM mixing pattern by a rotation $U_{12}$ from the left, possibly originating from the diagonalization of the charged lepton sector \cite{Frampton:2004ud,Romanino:2004ww,Altarelli:2004jb}
\be
U_{PMNS}=U_{12}(\alpha,\delta) U_{BM}=
\left(
\begin{array}{ccc}
\cos\alpha&e^{-i\delta}\sin\alpha&0\\
-e^{i\delta}\sin\alpha&\cos\alpha&0\\
0&0&1
\end{array}
\right) U_{BM}~~~.
\label{bimaxmod}
\ee 
To first order in $\alpha$ we have
\bea
\sin^2\theta_{12}&=&\frac{1}{2}+\frac{1}{\sqrt{2}}\alpha\cos\delta\nonumber\\
\sin\theta_{13}&=&\frac{1}{\sqrt{2}}\alpha\label{bm}\\
\delta_{CP}&=&\delta\nonumber\\
\sin^2\theta_{23}&=&\frac{1}{2}\nonumber
\eea
By eliminating $(\alpha,\delta)$ we get a relation between $\sin^2\theta_{12}$, $\sin\theta_{13}$, and $\delta_{CP}$, plotted in fig. \ref{figctm}. This model predicts $\delta_{CP}$ close to $\pi$
in order to reproduce correctly  $\sin^2\theta_{12}$, as can be seen from eqs. (\ref{bm}).

Another possibility is to relax the symmetry requirements. $S_4$ is the smallest group reproducing TB mixing through the breaking down to
$G_e=Z_3$ and $G_\nu=Z_2\times Z_2$, whose generators are $T$ and $(S,U)$, respectively
\footnote{It is well-known that, in concrete models, the TB mixing pattern can also be obtained from the group $A_4$, generated by $S$ and $T$, if $U$ arises accidentally due to a particular field content.}. 
In the basis where $T$ and the charged leptons are diagonal, the element $U$ coincides with the so-called $\mu\tau$ exchange symmetry \cite{Lam:2001fb,Kitabayashi:2002jd,Grimus:2003vx,Koide:2003rx,Ghosal:2003mq,Mohapatra:2005yu}, 
directly responsible for the vanishing of $\theta_{13}$ and for $\theta_{23}$ being maximal. We can avoid having $\theta_{13}=0$ and $\theta_{23}=\pi/4$ if
$G_\nu$ is a single $Z_2$ subgroup generated either by the element $S$ or by the element $SU$.
When the preserved parity is $S$, the mixing pattern, $TM_2$, is trimaximal and corresponds to 
\be
U_{TM_2}=U_{TB}~U_{13}(\alpha,\delta)=
U_{TB}~
\left(
\begin{array}{ccc}
\cos\alpha&0&e^{i\delta}\sin\alpha\\
0&1&0\\
-e^{-i\delta}\sin\alpha&0&\cos\alpha
\end{array}
\right)~~~,
\label{trimax}
\ee 
with both $\alpha$ and $\delta$ unconstrained. 
When the preserved parity is $SU$, the mixing pattern, $TM_1$, is also of trimaximal type and is given by $U_{TM_1}=U_{TB}~U_{23}(\alpha,\delta)$, where $U_{23}(\alpha,\delta)$
is the transformation analogous to $U_{13}(\alpha,\delta)$, acting in the 23 plane. The mixing angles and the Dirac phase are predicted in terms of $(\alpha,\delta)$ and we get two relations among
physical quantities, shown in Table \ref{tt3} \cite{He:2006qd,He:2011gb,Grimus:2008tt,Grimus:2009xw,Albright:2008rp,Antusch:2011ic,King:2011zj,Altarelli:2012bn}.
The interesting feature of these relations is that the predicted deviations from TB are linear in $\sin\theta_{13}$ for $\sin^2\theta_{23}$, and quadratic for $\sin^2\theta_{12}$, 
known with much better precision. One of these relations is plotted in fig. \ref{figctm} in the case of $TM_2$, from which we see that a substantial improvement in the data is needed
to test this possibility.
\begin{table}[h!]
\caption{\label{tt3} Sum rules for $TM_{1,2}$ mixing patterns.}
\begin{center}
\begin{tabular}{ll}
\br
$TM_1$& $TM_2$\\
\mr
$\sin^2\theta_{12}=\frac{1}{3}-\frac{2}{3}\sin^2\theta_{13}+O(\sin^4\theta_{13})$&$\sin^2\theta_{12}=\frac{1}{3}+\frac{1}{3}\sin^2\theta_{13}+O(\sin^4\theta_{13})$\\
$\sin^2\theta_{23}=\frac{1}{2}-\sqrt{2} \sin\theta_{13} \cos\delta_{CP}+O(\sin^2\theta_{13})$&$\sin^2\theta_{23}=\frac{1}{2}+\frac{1}{\sqrt{2}} \sin\theta_{13} \cos\delta_{CP}+O(\sin^2\theta_{13})$\\
\br
\end{tabular}
\end{center}
\end{table}
Explicit models based on $A_4$ realizing the $TM_2$ breaking pattern were indeed proposed before the measurement of $\theta_{13}$ \cite{Lin:2009bw}. The possibility of
reducing the residual symmetry $G_\nu$ to $Z_2$ can be systematically investigated \cite{Hernandez:2012ra}. 

\begin{figure}[h]
\center
\includegraphics[width=0.5\textwidth]{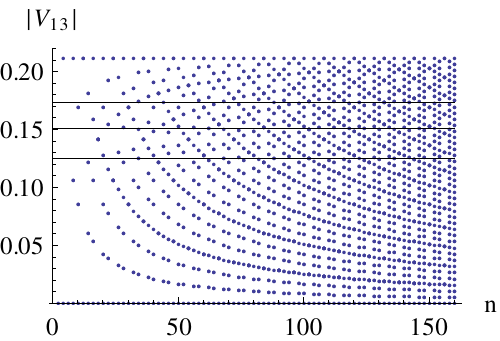}
\caption{Possible values of $|U_{e3}|$, indicated as $|V_{13}|$ in the vertical axis, versus $n$ in $\Delta(6 n^2)$ models, from ref.  \cite{King:2013vna}. The lines denote the present approximate $3\sigma$ range of $|U_{e3}|$. Examples include $|V_{13}|=0.211,0.170,0.160,0.154$ for $n=4,10,16,22$, respectively.  Each value of $U_{e3}$ allows for two values of $\theta_{23}$ with $\delta_{CP}=0$ and $\delta_{CP}=\pi$ given by $\theta_{23}=45^\circ \mp \theta_{13}/\sqrt{2}$ respectively.}
\label{u13vsna}
\end{figure}

A further possibility is to look for alternative LO approximations where $\theta_{13}$ is closer to the 
measured value. Remarkably, an infinite set of groups $G_f$ giving rise to LO approximations closer to the data has been found. Of particular interest is the special form of trimaximal mixing $TM_2$ in (\ref{trimax}),
with both $\alpha$ and $\delta$ quantized, reproduced by groups of the series $\Delta(6 n^2)$ \cite{Toorop:2011jn,deAdelhartToorop:2011re,Holthausen:2012wt,King:2013vna}. For example, choosing $n=(4,8,10)$ we have $\alpha=(\pm\pi/12,\pm \pi/24,\pm \pi/15)$ and $\sin^2\theta^0_{13}=(0.045,0.011,0.029)$. The Dirac phase is zero (modulo $\pi$). In fig. \ref{u13vsna} the values of $|U_{e3}|$ are
plotted versus $n$ \cite{King:2013vna}. 
Other discrete groups have been investigated in ref. \cite{Hagedorn:2013nra}.
Very remarkably, a complete classification of all possible mixing matrices $|U^0_{PMNS}|$ generated from any finite group has been recently carried out in ref. \cite{Fonseca:2014koa}.

Another development consists in combining discrete and $CP$ symmetries \cite{Feruglio:2012cw,Holthausen:2012dk} and exploring the symmetry breaking patterns such a combination can give rise to. 
A well-known example is that of the so-called $\mu\tau$ reflection symmetry \cite{mt1,mt2,mt3,mutaureflection_GL} (not to be confused with the $\mu\tau$ exchange symmetry), which exchanges a muon (tau) neutrino with a tau (muon) antineutrino in the charged lepton mass basis. 
If such a symmetry is imposed, the atmospheric mixing angle is predicted to be maximal, while $\theta_{13}$ is in general non-vanishing
for a maximal Dirac phase $\delta$ and the Majorana phases vanish. The solar mixing angle remains unconstrained.

A general formalism which combines $CP$ and flavour symmetries \cite{Feruglio:2012cw} can be used to constrain the lepton mixing matrix.
A theory symmetric under $CP$ and under a discrete flavour group $G_f$ is assumed to have residual symmetries $G_e$, generated by some elements $Q_i$ and $G_\nu=Z_2\times CP$, generated by
a parity $Z$ and a $CP$ transformation $X$. The action of $X$ in flavour space can be non-trivial \cite{Branco:2011zb} and should respect a set of consistency conditions \cite{Feruglio:2012cw,GCPV1,GCPV2,CPGf_HD}.
The residual symmetries $G_e$ and $G_\nu$ imply the following conditions on $m^\dagger_l m_l$ and $m_\nu$:
\be
Q^\dagger_i (m^\dagger_l m_l) Q_i=(m^\dagger_l m_l)~~~,~~~~~~~Z^Tm_\nu Z=m_\nu~~~,~~~~~~~X m_\nu X=m^*_\nu~~~.
\ee
These conditions are strong enough to determine $U^0_{PMNS}$ completely, up to one real parameter $\theta$, ranging from $0$ to $\pi$:
\be
U^0_{PMNS}=U^0_{PMNS}(Q_i,Z,X,\theta)~~~~~~~~~~~~~0\le\theta\le\pi~~~.
\ee
\begin{figure}
\begin{center}
\includegraphics{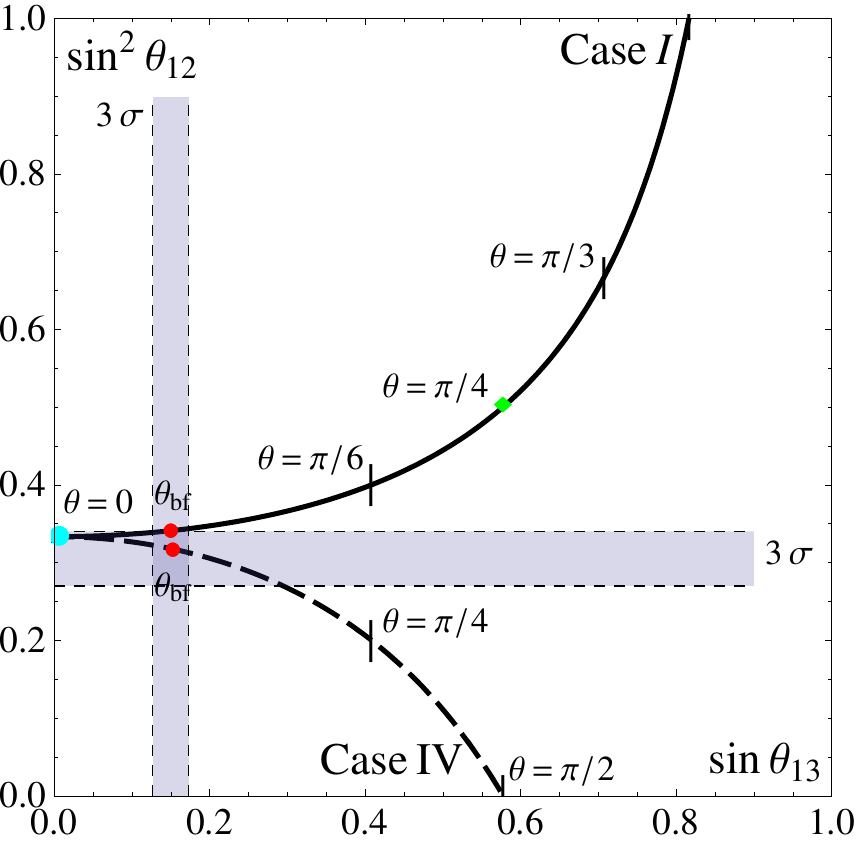}
\end{center}
\caption{\label{t4} Results for the mixing parameters $\sin \theta_{13}$, $\sin^2 \theta_{12}$  for Case I (straight line) and Case IV (dashed line), from ref. \cite{Feruglio:2012cw}. We mark the 
 value $\theta_{\mbox{bf}}$ of the parameter $\theta$ for which the $\chi^2$ functions have a global minimum with a red dot. $3 \, \sigma$ ranges for the mixing angles are also shown}
\label{cp}
\end{figure}
Mixing angles and phases, both Dirac and Majorana, are then predicted as a function of $\theta$, modulo the ambiguity related to the freedom of permuting rows and columns
and to the intrinsic parity of neutrinos. The formalism is completely invariant under any change of basis in field space. The physical results only depend on the initial symmetry
and the residual symmetries specified by $(Q_i,Z,X)$. An interesting example is provided by $G_f=S_4$. An exhaustive analysis has been presented in ref. \cite{Feruglio:2012cw}.
The residual symmetries can be chosen as $G_e=Z_3$, generated by the element $T$, and $G_\nu=Z_2\times CP$, generated by $(Z,X)$. 
The parity transformation $Z$ can be either $S$ (case I) or $SU$ (case IV) and a consistent $CP$ transformation $X$ acting on the lepton doublets  
coincides with the $\mu-\tau$ reflection symmetry in the basis where $T$ (and the combination $m^\dagger_l m_l$) is diagonal. 
Thus the predicted mixing pattern has a maximal atmospheric mixing angle, a maximal Dirac phase, vanishing Majorana phases and there is a relation
between the solar angle and the reactor angle, shown in fig. \ref{cp}.

Recently several explicit models combining $CP$ and flavour symmetries have been proposed and several series of discrete groups have been investigated in combination with $CP$ \cite{Feruglio:2013hia,Ding:2013hpa,Ding:2013bpa,Ding:2013nsa,King:2014rwa,Hagedorn:2014wha}.
Other approaches making use of CP and discrete symmetries are described in \cite{invariant_matrix_elements,Gupta:2011ct,simplest_nu_mass_S4,S4andCP_MN,gCPV1,gCPV2,gCPV3,gCPV4,gCPV5,CP_acc1,CP_acc2,CPV_Tprime1,CPV_Tprime2}.
\section{Hints from empirical relations}
Empirical relations among fermion masses and/or mixing angles have been frequently suggested as a clue towards a solution of the flavour puzzle.
Here as an example I will comment one of the most striking ones, Koide's formula for charged lepton masses \cite{Koide:1983qe}:
\be
\frac{\left(\sqrt{m_e}+\sqrt{m_\mu}+\sqrt{m_\tau}\right)}{\sqrt{m_e+m_\mu+m_\tau}}=\sqrt{\dd\frac{3}{2}}~~~~.
\label{koide}
\ee
The experimental values of the pole masses $m_e$, $m_\mu$, $m_\tau$ are \cite{Agashe:2014kda}:
\bea
m_e&=&0.510998928(11)~{\rm MeV}\nn\\
m_\mu&=&105.6583715(35)~{\rm MeV}\nn\\
m_\tau&=&1776.82(16)~{\rm MeV}~~~~.
\label{polemasses}
\eea
By using as input $m_e$ and $m_\mu$ in eq. (\ref{polemasses}), from Koide formula we find $m_\tau=1776.97$ MeV, with no appreciable dependence on the
experimental errors of $m_e$ and $m_\mu$ and in perfect agreement with the measured value.
There are several attractive features of Koide formula, beyond its simplicity. It is independent of the choice of mass units. It can be expressed as $K(m_e,m_\mu,m_\tau)=0$, where $K$ is an 
homogeneous function of the charged lepton masses, symmetric under permutations of $m_e$, $m_\mu$, $m_\tau$.
This makes the formula particularly attractive and has stimulated lot of interest and of activity aimed at deriving or embedding it within a more
fundamental framework. An unsatisfactory feature of the formula is that it requires as inputs pole masses and not running masses, which would be more adequate if masses are believed to
originate from a common scale $\Lambda_f$.
\begin{table}[h!]
\caption{\label{tx} Values of the running SM Yukawa couplings for the charged fermions in the $\overline{MS}$ scheme, at selected renormalisation scales $\mu$, from ref. \cite{Antusch:2013jca}.
The fourth and the fifth rows show the prediction of the mass relations in eqs. (\ref{koide}) and (\ref{new}), respectively, using as inputs the first two rows, $y_e(\mu)$ and $y_\mu(\mu)$.
The errors indicated in brackets affect the last digit and are dominated by the uncertainty in $y_\mu(\mu)$.}
\vskip 0.2cm
\begin{center}
\begin{tabular}{ c l l l l }
\hline
\noalign{\smallskip}
&{$\mu = M_Z$} & {$\mu= 1$ TeV} &  $\mu= 3$ TeV & {$\mu= 10$ TeV} \\
\noalign{\smallskip}
\hline
\noalign{\smallskip}
\hline
\noalign{\smallskip}
$y_e \;/\, 10^{-6}$ & $2.794745^{+0.000015}_{-0.000016}$ & $2.8482^{+0.0022}_{-0.0021}$ & $2.8646^{+0.0032}_{-0.0029}$ & $2.8782^{+0.0042}_{-0.0039}$\\
\noalign{\smallskip}
\hline
\noalign{\smallskip}
$y_\mu \;/\, 10^{-4}$ & $5.899863^{+0.000019}_{-0.000018}$ & $6.0127^{+0.0047}_{-0.0044}$ & $6.0473^{+0.0067}_{-0.0062}$ & $6.0761^{+0.0088}_{-0.0082}$ \\
\noalign{\smallskip}
\hline
\noalign{\smallskip}
$y_\tau \;/\, 10^{-2}$ & $1.002950^{+0.000090}_{-0.000091}$ & $1.02213^{+0.00078}_{-0.00077}$ & $1.0280\pm 0.0011$ & $1.0329^{+0.0014}_{-0.0015}$ \\[0.5ex]
\noalign{\smallskip}
 \hline
\noalign{\smallskip}
$y^K_\tau \;/\, 10^{-2}$ & $0.990448(3)$ & $1.0094(7)$& $1.015(1)$& $1.020(1)$\\
\noalign{\smallskip}
 \hline
\noalign{\smallskip}
$y_\tau^{(\ref{new})} \;/\, 10^{-2}$ & $0.991610(3)$ & $1.0106(8)$& $1.016(1)$& $1.021(1)$\\
\noalign{\smallskip}
\hline
\end{tabular}
\end{center}
\end{table}
This can be seen in table \ref{tx} where the running Yukawa parameters of the charged leptons in the $\overline{MS}$ scheme are listed for several values of the scale $\mu$.
The fourth row shows the value of the $\tau$ Yukawa coupling $y^K_\tau(\mu)$ as derived from the Koide formula using as inputs $y_e(\mu)$ and $y_\mu(\mu)$.
In the range $\mu=0.1 \div 10$ TeV the mismatch between $y^K_\tau(\mu)$ and $y_\tau(\mu)$ is close to one percent, while the accuracy with which 
$y_\tau(\mu)$ is estimated in ref. \cite{Antusch:2013jca} ranges from $10^{-4}$ to $10^{-3}$, thus making the discrepancy significant. For example the value of the tau mass at the scale  
 $M_Z$ predicted by the Koide relation is smaller by about 20 MeV, while the experimental precision on the same parameter is close to 0.2 MeV. 
This gap mainly arises as the effect of the pure QED running going from $m_e$ to $m_\tau$ as can be seen from the leading order relation between pole masses 
and $\overline{MS}$ masses:
\be
m_l(\mu)=m_l\left[1-\frac{\alpha}{\pi}\left(1+\frac{3}{2}\log\frac{\mu}{m_l}\right)\right]~~~~.
\ee
At energies higher than $1\div 10$ TeV, the success of the Koide relation depends on unknown physics. If the SM merges into a supersymmetric theory, the evolution of the Yukawa couplings depends on additional parameters, like the superparticle thresholds and $\tan\beta$. Just above the superpartners
mass threshold the Yukawa coupling are known to a much worst precision, around one percent \cite{Antusch:2013jca}, and the Koide relation might again be compatible with the extrapolated data.
At the GUT scale, larger uncertainties are induced on $y_{e,\mu,\tau}(M_{GUT})$ by $\tan\beta$.

More in general, we can ask what is the probability of finding a simple homogeneous relation among charged fermion masses holding to an accuracy similar to the one of the Koide relation 
at a given scale $\mu$. It is difficult to answer quantitatively this question, but I personally think that such probability is relatively high. As an example
consider the following relation, valid for charged fermion masses rather than for their squared roots:
\be
\left\vert
\frac{\omega m_e+\omega^2 m_\mu+m_\tau}{m_e+m_\mu+m_\tau}
\right\vert
=\dd\frac{11}{12}~~~~~~~~~~~\omega=e^{\dd i\frac{2\pi}{3}}~~~.
\label{new}
\ee
This relation, symmetric under permutations of the flavour labels, produces the outputs in the fifth row of table \ref{tx}. We see that the predictions of $y_\tau(\mu)$ are as good as the one from the Koide relation.
Probably it is not surprising that scanning thousands of possibilities we can find ``simple'' relations working at the level of the percent accuracy. 
Much more difficult is to find, through these relations, a direct link to some unknown fundamental layer of particle physics.
\section*{Conclusion}
We are witnessing a continuous experimental progress in flavour physics. In neutrino physics, squared-mass differences and mixing angles are known to an accuracy that approaches the percent level.
The reactor angle $\theta_{13}$ is away from zero by many standard deviations.
For the first time global fits hint at a non-trivial Dirac phase. 
While the new data have been effective in ruling out many models of fermion masses
and mixing angles, as a matter of fact no compelling and unique theoretical picture has emerged so far. 

Present data can still be described within widely different frameworks. Based on our experience with gauge interactions we might hope that
the flavour sector becomes simple and symmetric at a high energy scale, with a small number of relevant parameters providing a complete description.
It is fair to say that we have not been able to identify a clear symmetry pattern from data so far. Before 2012 discrete symmetries were considered as
a good candidate. In particular those predicting a nearly tri-bimaximal mixing were favored by data, but the prediction of $\theta_{13}$ turned out to be wrong. 
The evidence for discrete symmetries in the quark sector is very poor
and in a unified description of all fermions this kind of symmetry is typically badly broken in the quark sector. The whole approach is too much centered on the lepton mixing properties,
while a description of the fermion masses seems to need additional ingredients. Several modifications of the simplest schemes to accommodate the present data are still 
possible and have the advantage of being quantitatively testable. But the real open question is whether a non-trivial implementation of discrete symmetries exists encompassing quark and lepton sectors in a unified picture and providing a quantitative description of both masses and mixing angles.

There are other models where fermion masses and mixing angles are mapped into a large number of irreducible and unconstrained order-one parameters, thus incarnating the
Anarchy idea. 
For their intrinsic nature these models essentially escape experimental tests going beyond the order-of-magnitude accuracy.
However we cannot fail to be impressed by the fact that they can provide a common description to both fermion masses and mixing angles, that they are compatible
with grand unified theories and that they can be derived within widely different theoretical frameworks. The fact that this kind of models
can be implemented even in a highly constrained setup such as an SO(10) grand unified theory is really remarkable. 
As a drawback, in these models  the bounds on the scale of new flavored physics is typically pushed above the 10 TeV scale,
reducing the possibility of testing these ideas.

Flavour remains a fascinating mystery, still eluding all our attempts to find the rationale underlying our observations. Has this puzzle any solution? 
Are we misled by the questions we have formulated so far? Will we ever have access to the flavour scale? Man has long been 
fascinated by the mystery of planet motion. Surprisingly precise measurements and accurate predictions already existed in remote epochs.
For a long time the most reliable models were based on the special character of geometrical figures like the sphere or the circle.
Attempts to explain the relative sizes of the solar planetary orbits revealed themselves misleading. More accurate observations, perseverance in identifying the correct questions
and renunciation of old prejudices opened new perspectives to the scientific thought. Will this be the fate of the flavour puzzle too?\vspace{1.cm}
\subsection*{Acknowledgments} 
I would like to thank warmly Guido Altarelli, Reinier de Adelhart Toorop, Belen Gavela, Walter Grimus, Manmohan Gupta, Claudia Hagedorn, Isabella Masina, Luciano Maiani, Luca Merlo, Paride Paradisi, Massimo Passera, Ketan Patel, Stefano Rigolin, Andrea Romanino, Denise Vicino, 
Robert Ziegler for useful correspondence, discussions, observations and the pleasant collaborations  on which this review is based. I thank the Institute for Theoretical Physics (IFT) UAM-CSIC in Madrid for hospitality while preparing part of this review.
This work was supported in part by the MIUR-PRIN project 2010YJ2NYW and by the European Union network FP7 ITN INVISIBLES (Marie Curie Actions, PITN-GA-2011-289442).
\section*{References}

\end{document}